# Structural properties of one-dimensional $Cs_2CoCl_4$ confined within single-walled carbon nanotubes

Jaskaran S. Mangat,[1] Yu Lei,[2] Matthew Weyland,[3,4] Yisong Han,[2] Kiran Bal,[2] Martin R. Lees,[2] Craig I. Hiley,[1] Piotr Dłużewski,[5] Sławomir Kret,[5] Peng Wang,[2] Richard I. Walton[1] and Jeremy Sloan[2,*]

[1]*Department of Chemistry, University of Warwick, Gibbet Hill Road, Coventry, CV4 7AL, United Kingdom.*
[2]*Department of Physics, University of Warwick, Gibbet Hill Road, Coventry, CV4 7AL, United Kingdom.*
[3]*Monash Centre for Electron Microscopy, Monash University, Clayton, Victoria, 3800, Australia.*
[4]*Department of Materials Science and Engineering, Monash University, Clayton, Victoria, 3800, Australia.*
[5]*Instytut Fizyki PAN, Al. Lotników 32/46, 02-668 Warszawa, Poland.*

Crystals under one-dimensional (1D) confinement are well-known to exhibit drastic changes in metallicity, magnetic properties and chemical state, however, the intermediate phase space between binary metal halides and ternary metal halide perovskites remains poorly explored, especially in the context of the rich polymorphism exhibited by both families in the one-dimensional limit. Through aberration-corrected (scanning) transmission electron microscopy and multislice simulations, it is shown that the metal halide $Cs_2CoCl_4$ crystallizes in the tetragonal $\wp 4/mcc$ and orthorhombic $\wp mcm$ rod groups under radial compression within single-walled carbon nanotubes (SWCNTs) of increasingly small diameter, with a massive re-entrant orthorhombic strain towards the 1 nm extremum. The persistence of $Co^{2+}$ is determined from fits to the d.c. magnetization, with a surprisingly small increase in the effective moment (4.607(3) to 4.788(3) $\mu_B$/f.u.) and Weiss constant (−7.9(3) to −4.09(7) K) after confinement in the SWCNTs, suggesting that the confined structure topologically preserves the core magnetic properties of the bulk. Both unconventional polymorphs observed are noticeably different to the high-pressure piezochromic polymorph previously shown to undergo a tetrahedral-to-octahedral coordination transition, highlighting 1D confinement as a unique tool for structural manipulation.

## I. INTRODUCTION

Single-walled carbon nanotubes (SWCNTs) are renowned for their exceptional physical properties, including terapascal Young's moduli arising from strong $sp^2$ orbital hybridization [1,2], ballistic electron transport due to gapless nodes in the band structure [3,4], and variable metallicity dependent on the nanotube chirality [5,6]. Through chemical diffusion and melt insertion [7], true one-dimensional (1D) atomic chains [8,9], nanoribbons [10,11] and more complex structures such as clusters [12,13] and nanohelices [14] can be stabilized within the SWCNTs under extreme steric strain, causing various unusual properties to emerge including charge density waves [15], torsional instabilities [16], changes in oxidation state [17] and Tomonaga-Luttinger liquidity [18].

Most recently, new families of halide perovskites with complex chemistries such as $Cs_3PbBr_5$ and $Cs_4Sn_4I_{12}$ have been shown to consistently form in ~1.4 nm wide SWCNTs with widened bandgaps [19,20], with even high-entropy metal halides containing various rare earth ions crystallizing uniformly within the SWCNTs up to a senary composition [21]. Excitingly, these confined materials exhibit a highly tuneable photoresponse in heterojunctions for photodetectors [22,23] and a subthreshold swing as low as 35 mV/dec in field-effect transistors [24], ideal for modular dual-band sensitivity and ultralow power operation in optoelectronic devices. The nanotube sheath also offers a protective barrier against water vapour and polar solvents, allowing devices to operate under ambient conditions even for notoriously air-sensitive bromide and iodide-based compositions [20,21]. Binary metal halides with part-filled *d*-bands have also been observed to undergo drastic changes in magnetic ordering under 1D confinement, with a prominent example being chromium chloride: bulk $CrCl_3$ hosts edge-shared octahedra and is antiferromagnetic at low temperatures [25], becoming a ferromagnet when reduced to a single monolayer due to the elimination of the out-of-plane antiferromagnetic exchange [26]. However, $CrCl_3$ becomes a spin glass within 1 – 2 nm wide SWCNTs below 3 K as evidenced by a large frequency-dependent change in the imaginary component of the a.c. susceptibility [27], with a separate study reporting a structural rearrangement to *face*-shared octahedra in ~1.1 nm wide SWCNTs and suggesting that charge transfer from the SWCNTs turns the system into a ferromagnet [28]. Further complicating the issue, the stoichiometry changes to $CrCl_2$ ($Cr^{2+}$) with square-planar coordination in 0.92 nm wide SWCNTs [17]. Although there is a qualitative increase in the structural symmetry of these metal halides with shrinking SWCNT diameter, the exact symmetries and nature of the magnetostructural coupling are unclear, exacerbated by some studies failing to systematically analyse *Z*-contrast of annular dark-field images and sample heterogeneity often inhibiting detailed quantitation of the magnetization [21,24,27]. Similarly to monolayer $CrCl_3$, $Cs_2CoCl_4$ at low temperatures is an antiferromagnetic realization of the *XY* model in the 3*d* metal halides [29,30], but under ~7 GPa of pressure a

Corresponding author:
*j.sloan@warwick.ac.uk

piezochromic structural phase transition occurs, accompanied by a tetrahedral-to-*octahedral* change in coordination [31]. Thus, confining $Cs_2CoCl_4$ within SWCNTs could provide a link between binary metal halides with competing magnetic interactions and ternary metal halide perovskites, particularly in the context of magnetostructural coupling and SWCNT diameter-dependent crystallization in the vast unknown 1D phase space. Motivated by this prospect, we synthesized $Cs_2CoCl_4$@SWCNT *via* melt insertion and studied the structural polymorphism in detail through annular dark-field scanning transmission electron microscopy (ADF-STEM), high-resolution transmission electron microscopy (HRTEM) and multislice simulations, with extensive knowledge of our sample constraints allowing us to extract various quantities including the global filling ratio from the d.c. magnetization.

## II. METHODS

The precursor material $Cs_2CoCl_4$ was prepared hydrothermally with orthophosphoric acid as described similarly by Jiang *et al.* [32]. Compositional homogeneity was verified by powder X-ray diffraction at Diamond Light Source Beamline I11, and a single crystal from the same batch was analysed at room temperature on a Rigaku Oxford Diffraction XtaLAB Synergy-S single-crystal diffractometer using Mo Kα radiation in order to estimate values of isotropic atomic displacement parameters $U_{iso}$ for the nanostructures in our multislice simulations. Structure solution was implemented with SHELXT [33] and refinement with the Olex2.refine algorithm [34,35]. Further information is given in Fig. S1 and Table S1 in the Supplementary Information.

NanoIntegris® PureTubes™ SWCNTs were purchased from Merck with a reported diameter range of 1.2 – 1.7 nm and metal catalyst content of <0.34 wt% from the corresponding data sheet [36]. The orientation of the SWCNT bundles was randomized by dispersal in chloroform before microfiltration. Afterwards, carbonaceous impurities were removed by oxidation in air at 440 ºC for 30 minutes, with the SWCNTs subsequently being thoroughly ground with the $Cs_2CoCl_4$ precursor in a ~1:1 volume ratio and transferred to an ampoule under $N_2$ atmosphere (sealed under dynamic vacuum) before holding at 615 ºC for two days to allow the molten precursor to percolate into the SWCNTs. The ampoule was brought back to room temperature at a rate of 1 °C/min, with extraneous crystallites and ferromagnetic metal catalyst subsequently washed away (corroborated by a reduction in the high-temperature susceptibility after washing, see Fig. S2) by stirring in 37% HCl aqueous solution for 30 minutes at room temperature, before microfiltration and drying at 70 ºC to obtain the sample analysed in this study.

ADF and bright-field (BF) STEM imaging were performed on a JEOL ARM200F microscope equipped with a Schottky field emission gun, and HRTEM imaging on spherical aberration-corrected FEI Tecnai and Thermo Scientific Spectra $\varphi$ microscopes. Images were acquired at Scherzer defocus with an accelerating voltage of 80 kV to avoid knock-on damage from prolonged irradiation [37-39], with the Spectra $\varphi$ using an excited monochromator to minimize chromatic aberration. Energy-dispersive X-ray spectroscopy (EDX) scans were taken in quick succession in 22 separate regions at 200 kV on the ARM200F using an Oxford Instruments X-Max 80 detector. Image analysis was done in DigitalMicrograph™ with Fourier filtering and a mild gamma filter applied to enhance contrast. The $Z$-contrast *i.e.* partial atomic number dependence of the incoherent scattering intensity is well-established in ADF imaging [40-43], and was used for elemental fingerprinting to help devise the structural models. Models of the nanostructures were created in CrystalMaker® 11 [44], with the nanotubes imported from the software Nanotube Modeler v1.8.0 [45]. Multislice simulations of the model structures were implemented in clTEM v0.3.4 [46] and SimulaTEM [47] with an accelerating voltage of 80 kV for comparison with the ADF-STEM and HRTEM images respectively. The clTEM simulations consider thermal diffuse scattering using a frozen phonon method, wherein displacements are approximated by a Gaussian function based on $U_{iso}$ values provided by the user [48]. $U_{iso} = 0.006$ Å$^2$ was fixed for all C atoms with values for the other atoms tabulated in Table S1. Inner and outer detector radii were set to 45 and 180 mrad, with projected potentials calculated for 1 Å slices and a Gaussian blur filter applied with the kernel's standard deviation set up to 16 for the simulations [49].

Magnetization measurements were performed on a Quantum Design MPMS-5S magnetometer with ~10 mg of powdered sample in gelatin capsules. The field-dependent d.c. magnetization at 5 K was fitted to a Brillouin function to extract the molar mass $M_r$ per magnetic ion and the effective angular momentum quantum number $J$ [50], as in Equations (1) and (2):

$$M = \frac{mN_A g\mu_B J \mathfrak{B}_J}{M_r} \quad (1)$$

$$\mathfrak{B}_J = \frac{2J+1}{2J}\coth\left(\frac{g\mu_B H(2J+1)}{2k_B T}\right) - \frac{1}{2J}\coth\left(\frac{g\mu_B H}{2k_B T}\right) \quad (2)$$

where $m$ is the mass of the sample, $M$ is the measured magnetization, $N_A$ is the Avogadro constant, $k_B$ is the Boltzmann constant, $\mu_B$ is the Bohr magneton, $H$ is the applied magnetic field, $\mathfrak{B}_J$ is the Brillouin function and $g$ is the Landé $g$-factor (value determined as 1.986 from electron paramagnetic resonance). The value of the molar mass is complicated by the filling ratio of the SWCNTs, but it is known that the 1.2 – 1.7 nm average diameter range of NanoIntegris® PureTubes™ SWCNTs in the sample maps onto an approximate range of (9,9) – (13,13) chiral indices $(n,m)$ where $n, m \in \mathbb{Z}$, thereby mapping onto a similar nanotube width range of 1.22 – 1.76 nm (since $d = \frac{a\sqrt{3}}{\pi}\sqrt{n^2 + m^2 + nm}$, where $d$ is the SWCNT diameter and $a$ is the ~0.142 nm bond length in the hexagonal graphene lattice [5]) and corresponding to $C_{54}$ – $C_{78}$ (*i.e.* $C_{66\pm12}$) per repeating unit of the $Cs_2CoCl_4$ nanostructures observed. Therefore, scaling the number of C atoms per Co in the assumed $Cs_2CoCl_4C_{(66\pm12)x}$ global chemical formula by a factor $x$ allows the filling ratio $(= 1/x)$ to be calculated from the refined molar mass. This method does not account for the presence of any remaining extraneous bulk crystallites, but the global ratio of Cs:Co:Cl in the sample must be the same in order to conserve the precursor's stoichiometry (additionally determined by our EDX analysis in Fig. S3), justifying this approximation. Using the number of moles refined (= $m/M_r$), it was possible to fit a modified Curie-Weiss law to the temperature-dependent d.c. molar magnetization in order to extract the mean effective moment [51], with a negative temperature-independent term $\chi_0$ originating from the orbital diamagnetism of the SWCNTs themselves (on the order of $10^{-4}$ emu·mol$^{-1}$·Oe$^{-1}$ [52]). Such a temperature-independent term was not needed to fit the susceptibility of the polycrystalline bulk precursor in Fig. S2(a).

## III. RESULTS AND DISCUSSION

Bulk $Cs_2CoCl_4$ crystallizes in the orthorhombic *Pnma* space group, with lattice parameters of $a$ = 9.7776(2) Å, $b$ = 7.4005(1) Å and $c$ = 12.9850(2) Å in line with previous observations [53]. After obtaining the final sample, the most clearly resolvable crystal structures were examined in detail: generally this is in isolated SWCNTs away from thick bundles due to the minimization of background incoherent scattering from neighbouring regions. Firstly, a filled wide outlier SWCNT (~2.6 nm, approximated by (19,19) chiral indices) is observed in Fig. 1: the incoherent scattering is well-modelled by a supercell of the known tetragonal *I*4/*mcm* structure of $Cs_3CoCl_5$ [54], but with the radial limit of the SWCNT removing four columns of CsCl at the cross-sectional diagonal extrema and thereby necessitating a change in the stoichiometry. ADF and BF images of the surrounding environment in Figs. S4 (a, b) clearly show that the crystal is encapsulated within a wide SWCNT and is not an extraneous nanocrystal. Due to the lack of translational symmetry in two dimensions, the average group symmetry is better described by the subperiodic rod groups rather than space groups [55,56]: using the program RODS [57] with a penetration direction down [001] into the origin, the tetragonal rod group $\mathscr{p}4/mcc$ (full Hermann-Mauguin symbol $\mathscr{p}4/m2/c2/c$) is derived from the parent *I*4/*mcm* symmetry, with lattice parameters $a$ = $b$ ≈ 18.4 Å and $c$ ≈ 14.5 Å (with $c$ down the growth axis). The experimental and simulated line profiles in Fig. 1(d) are remarkably similar in both intensity and periodicity, although we note some blurred bleeding intensity of adjacent $Cs_\delta$ in-between the longitudinal atoms. Evaluating the unit cell, a non-integer stoichiometry of $Cs_{2.083}CoCl_{3.417}$ (approximately $Cs_2CoCl_{3.4}$) is obtained, which is different to both $Cs_2CoCl_4$ and $Cs_3CoCl_5$ due to surface termination of ions in the extremely small crystal. The charge neutrality condition nominally assigns an unconventional formal oxidation state of +1.33 to cobalt in this case, but two alternative scenarios are more plausible: additional chloride ions at the limit of detection of the microscope, or charge transfer between the SWCNT and the crystal; both acting to balance a $Co^{2+}$ valence state. Detailed inspection of Fig. 1(a) shows two faint spheres which we believe are Fourier artefacts from the thick SWCNT bundle rather than additional chloride ions due to the strong local Coulomb repulsion in those regions, with lone caesium and cobalt also being precluded due to the extremely weak scattering intensity and unfeasible local coordination respectively. Hence, we tentatively suggest that there is instead some degree of electron transfer (0.67 $e^-$ per unit cell) from the SWCNT to the crystal. This is plausible since it is known that $PbI_2$, AgBr and $CrX_3$ ($X$ = Cl, Br, I) act as electron acceptors within SWCNTs on the order of 0.02, 0.11 and 0.25 $e^-$ per unit cell respectively [28,58,59], although it is noted that other systems such as 1D elemental europium can act as electron *donors* for the nanotube up to 1.79 $e^-$ per atom [18]; however, accurate quantification of any charge transfer is impossible on a single example and is hence outside the scope of this study.

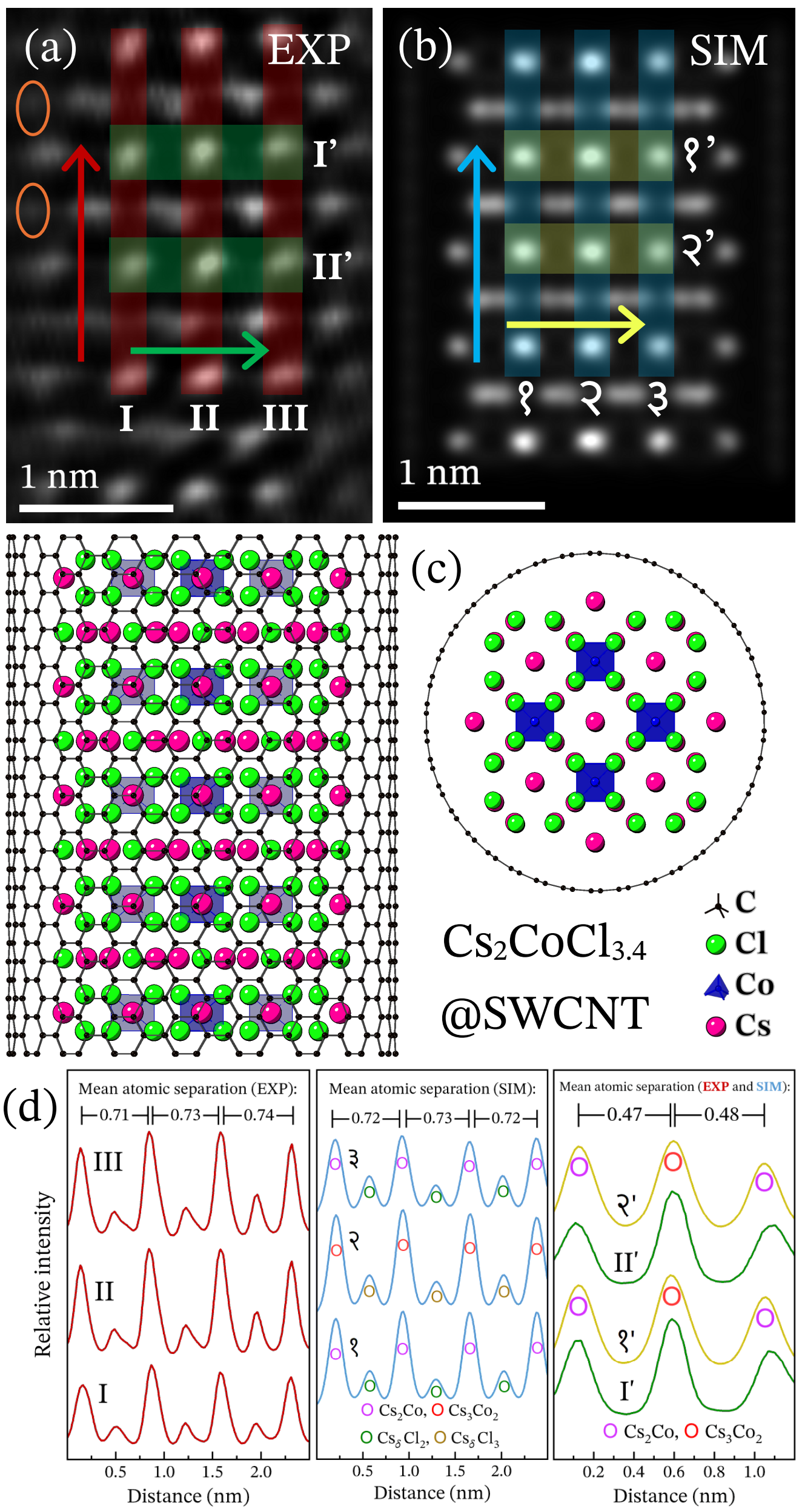

FIG. 1. (a, b) Fourier-filtered ADF-STEM image of a confined crystal within a wide SWCNT, with comparison to a frozen phonon simulated image of our structural model. Orange ovals highlight likely faint Fourier artefacts. (c) Model of the $\mathscr{p}4/mcc$ tetragonal structure within a (19,19) SWCNT ($d$ ~2.58 nm), hosting layered isolated $[CoCl_4]^{2-}$ tetrahedra. (d) Line profile scans across the experimental and simulated images, showing similar periodicities (within ±0.2 Å) and intensities in accordance with the ADF $I \sim Z^{1.4-1.9}$ law [43].

The structure of $Cs_2CoCl_4$ in 1.2-1.3 nm wide SWCNTs is next discussed, within the diameter distribution reported by NanoIntegris® [36]. The most commonly observed perspective of this structure is shown in Figs. 2, S4 (d, e) and S5 (d, e). It adopts a hexagonal motif of strongly scattering ions, with increased intensity in the inner two columns parallel to the growth axis. A model of $Cs^+$ ions with overlapping $[CoCl_4]^{2-}$ tetrahedra related by a $2_1/m$ screw operation excellently reproduces the observed ADF intensities in Fig. 2, specifically in the orthorhombic rod group *pmcm* (full symbol $p2/m2/c2_1/m$) with lattice parameters $a \approx 7.8$ Å, $b \approx 4.6$ Å and $c \approx 7.4$ Å, and the same stoichiometry as the bulk precursor. There is no known Cs-Co-Cl structure that adopts this or a similar structure, with the other unmentioned compositional derivative $CsCoCl_3$ crystallizing in the $P6_3/mmc$ space group with face-shared $[CoCl_6]^{4-}$ octahedra [60]. The HRTEM image of the same motif in Fig. 2(e) also possesses similar line scan periodicities to the corresponding simulation in Fig. 2(f), including the characteristic zigzag bleeding of chlorine intensity and even the broad silhouette of the tetrahedra in the inner columns, although the loss of strong $Z$-contrast causes a larger discrepancy between the observed and simulated latitudinal intensities compared to the ADF case. The latitudinal atomic separation appears to decrease slightly towards the bottom of Fig. 2(e), which may be explained by the SWCNT twisting or the crystal rotating slightly. The wall-to-wall measured SWCNT diameter is 1.23 nm as described in Fig. 2(g), but this is expected to be ~5% smaller than the real diameter (estimated at 1.29 nm) due to the well-known physical limitation that minima in the total projected electrostatic potential of the SWCNT do not coincide with the positions of the outermost carbon atoms [61]. A 60° rotation of the same unconventional polymorph is also observed in Fig. 3: the hexagonal motif transitions to an intense zigzag dominated by the Rutherford scattering of $Cs_2$, with faint points corresponding to $CoCl_2$ in the voids (an additional example of this perspective is given in Fig. S5(a)). The model in Fig. 3(c) reproduces the observed ADF intensities well, with the periodicities being equivalent to the longitudinal periodicities in Fig. 2(g) and the maximal difference between both motifs being ±0.1 Å. Even for the latitudinal scans in Fig. 2, the difference does not exceed ±0.5 Å (within the resolution of our microscopes), strongly supporting the proposed overall structural model. The orthorhombic strain, defined as $\varepsilon = (a-b)/(a+b) = 0.258$, is massive and testifies the stability of the ternary composition, which is especially interesting given that $\varepsilon$ for the structure reported in Fig. 1 is zero (*i.e.* tetragonal) and is lower in the bulk phase (0.139, although the bulk phase is less comparable since it has 3D translational periodicity; see Fig. S6 for a visual map of the structural relationships). To the best of our knowledge, such a large re-entrant orthorhombic instability on radial compression has not been reported in any other metal halides encapsulated within SWCNTs, nor is it plausibly explained by a deformed SWCNT cross-section. This is because the metastability of radial deformation is dependent on the interwall van der Waals forces being strong enough to overcome the bending strain energy for SWCNTs, which is only favourable above a critical diameter of 2.1 nm, well outside the general range of NanoIntegris® PureTubes™ SWCNTs [62].

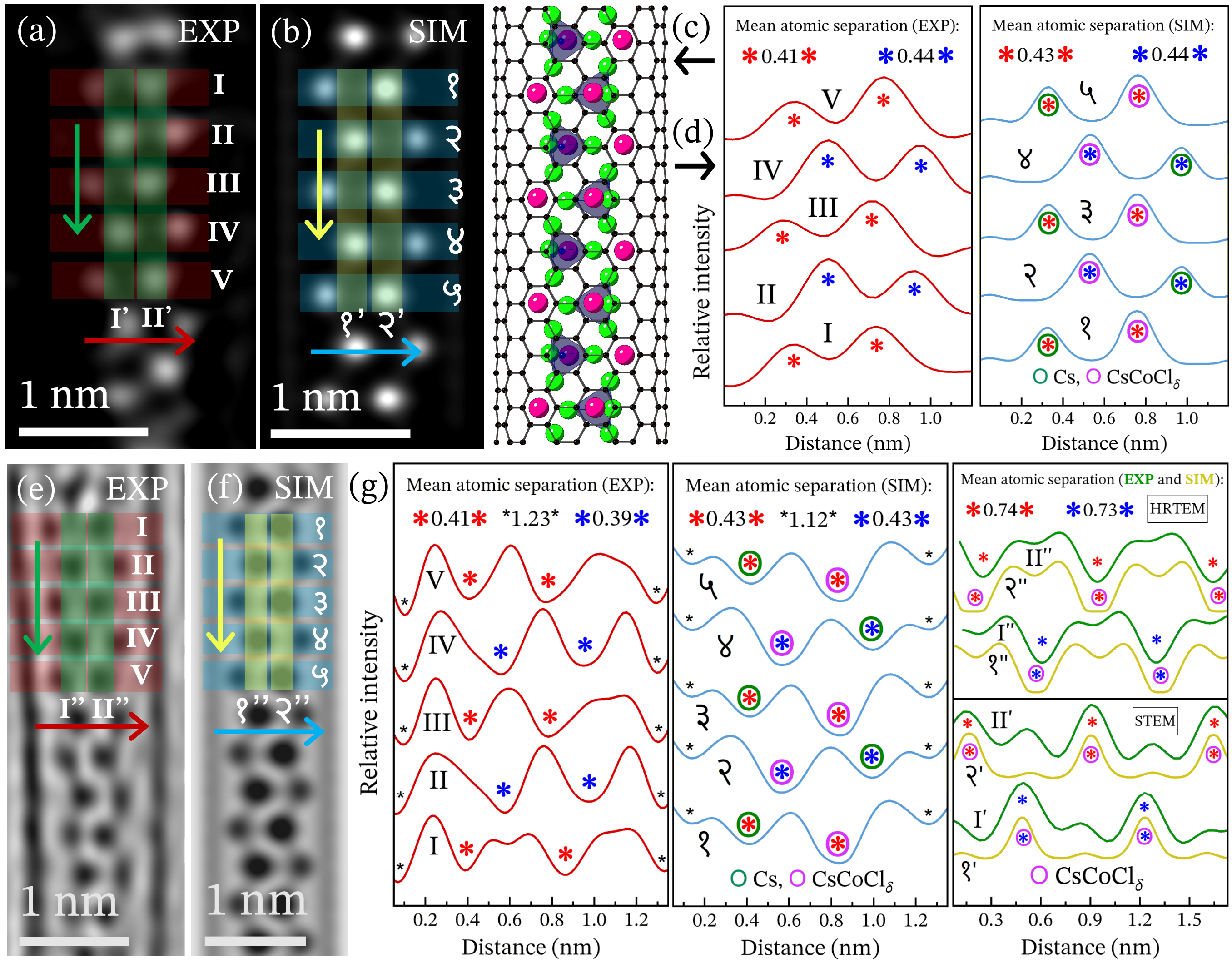

FIG. 2. (a, b) Experimental and simulated ADF images of $Cs_2CoCl_4$ in a hexagonal motif within a ~1.2-1.3 nm wide SWCNT. (c) Model of the imaged *pmcm* orthorhombic crystal structure of $Cs_2CoCl_4$, within a template (9,9) SWCNT. (d) Comparison of latitudinal line profile scans across the ADF-STEM images, highlighting the mean atomic separation and relative atomic intensities. (e, f) Experimental and simulated images of the same structure in HRTEM, with the SWCNT walls more easily visible. (g) Line profile scans across the HRTEM images: periodicities are within similar error to the ADF-STEM images, but with a loss of strong $Z$-contrast due to the phase interference of the exit wave dominating the HRTEM intensity. Longitudinal line profile scans across all images are also shown (peaks without asterisks are caused by bleeding intensity from neighbouring columns).

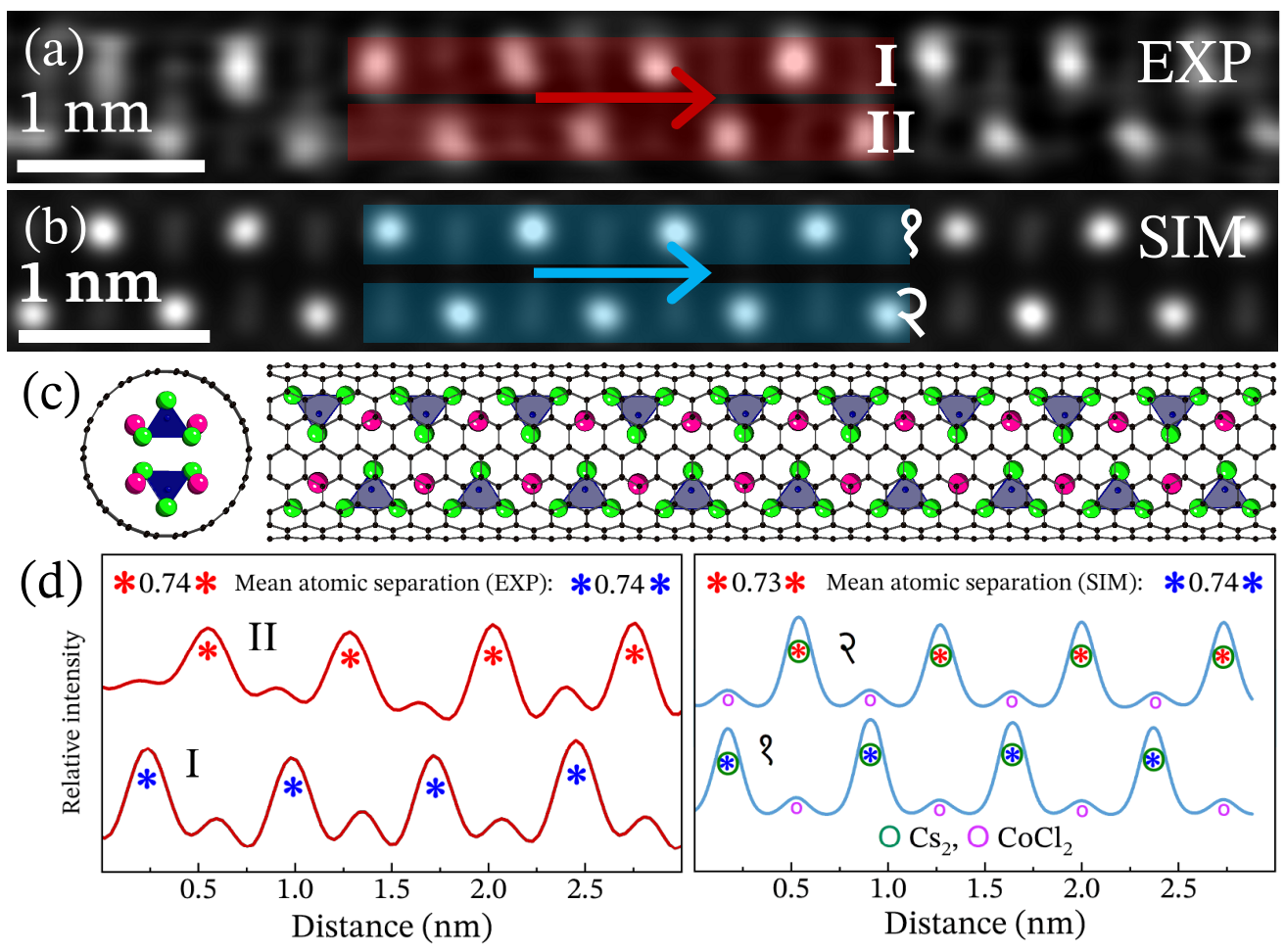

FIG. 3. (a, b) Experimental and simulated ADF images of orthorhombic $\mathcal{p}mcm$ $Cs_2CoCl_4$ in a zigzag motif. (c) Structural model with isolated $Co^{2+}$ tetrahedra on a triangular lattice. (d) Comparison of longitudinal line profile scans, with the experimental and simulated atomic periodicities being almost exactly equivalent (±0.1 Å).

Considering Neumann's principle [63], the implications of such a strong structural instability on electronic and magnetic renormalizations are of interest. Since this often manifests as a change in the 3*d* metal cation's oxidation state or crystal field in metal halides, *e.g.* the $Cr^{3+} \rightarrow Cr^{2+}$ valence transition in $CrCl_3$@SWCNT [27] or the octahedral crystal field restabilization in bulk $Cs_2CoCl_4$ under pressure [31], the possibility of a similar transition occurring here is examined. As seen in Fig. 4, analysis of the field-dependent d.c. magnetization up to 50 kOe using Eqs. (1) and (2) results in a best fitting $M_r$ of 1980(30) g·mol$^{-1}$ (approximately $C_{126}$ per unit cell) with an effective angular momentum quantum number of $J$ = 1.350(1). Owing to the often significant spin-orbit interaction in cobalt-based systems, the $J$ value is not readily interpretable as being reflective of the spin momentum, but it is qualitatively congruent with the expected $S$ = 3/2 spin state of $Co^{2+}$ in an octahedral or tetrahedral crystal field as depicted in Fig. 4(c). The filling ratio significantly affects the Brillouin fit and converges stably in any reasonable case, with the best fit calculated as 52(10)% within the approximate 1.22 – 1.76 nm SWCNT diameter range per unit cell, assuming that the $Cs_2CoCl_4$ stoichiometry homogeneously crystallizes within the SWCNTs. EDX in Fig. S3 shows that most confined structures indeed host this or a similar stoichiometry, however a minority are cobalt-enriched or cobalt-vacant. The intrinsic negative curvature in the inverse temperature-dependent susceptibility is well-modelled by a diamagnetic term $\chi_0$ in a modified Curie-Weiss law, determined as $-5.1(1) \times 10^{-4}$ emu·mol$^{-1}$·Oe$^{-1}$ if fitting to the whole field-cooled sweep. This is on the same order of magnitude as the experimentally reported diamagnetic susceptibility of purified carbon nanotubes (approximately $-1 \times 10^{-4}$ emu·mol$^{-1}$·Oe$^{-1}$ at room temperature [52]), but with a somewhat different value due to the additional diamagnetism of the confined $Cs_2CoCl_4$ crystals (being $-1.756 \times 10^{-4}$ emu·mol$^{-1}$·Oe$^{-1}$ from tabulated values [66]) and the possible presence of trace impurities after washing (*e.g.* as in Fig. S5(f)). The effective moment of 4.788(3) $\mu_B$/f.u. is consistent with $Co^{2+}$ in both a tetrahedral or octahedral crystal field and well out of the range of $Co^{3+}$, ruling out a reasonable valence transition and being slightly higher than the effective moment of bulk $Cs_2CoCl_4$ (4.607(3) $\mu_B$/f.u. as given in Fig. S2(a)).

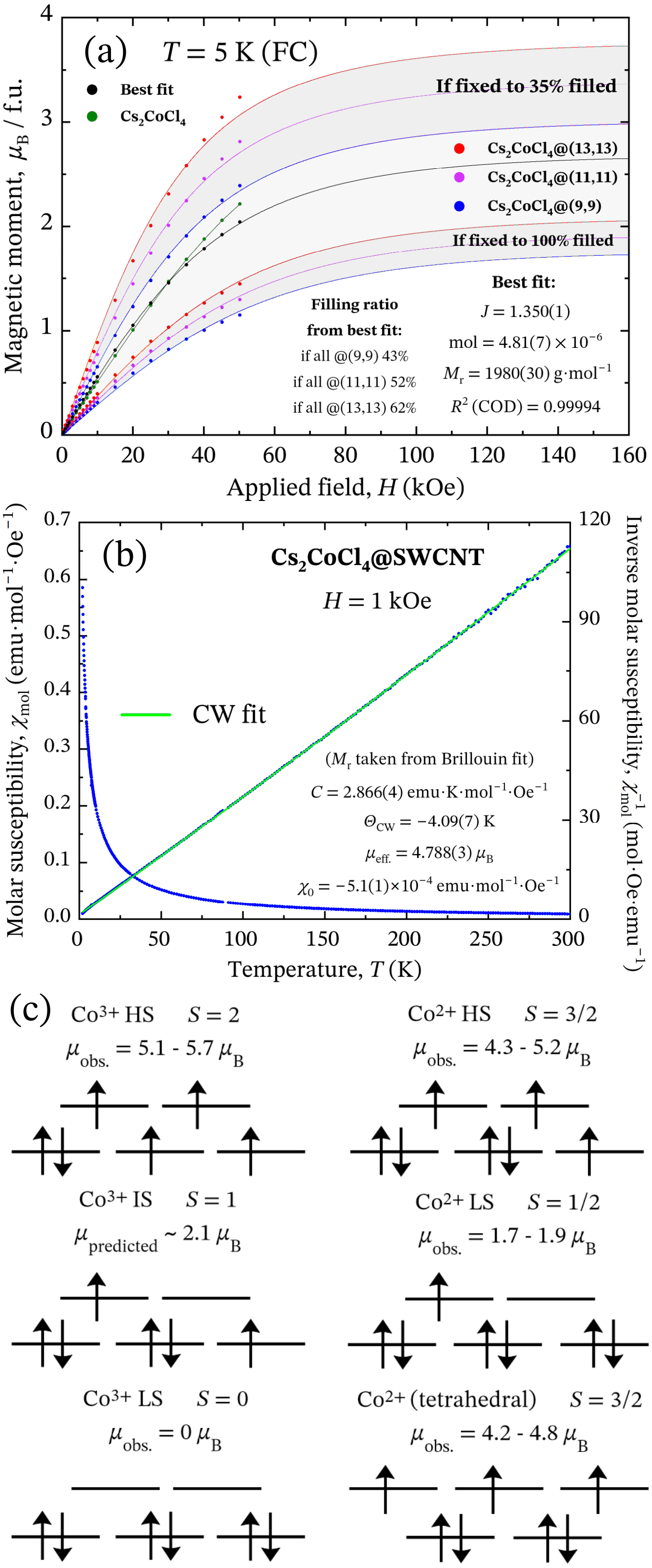

FIG. 4. (a) Brillouin fits to the field-cooled (FC) d.c. magnetization of $Cs_2CoCl_4$@SWCNT measured up to 50 kOe at 5 K. The fitted curves of two arbitrary fixed filling ratios in different fixed bounds of the SWCNT average diameter range (1.2 – 1.7 nm approximated by $C_{(66\pm12)x}$ per unit cell) are shown with comparison to the best fitting curve to visually describe the effect of both parameters, both ultimately affecting the $M_r$ refined. (b) Curie-Weiss fit to the FC susceptibility using the $M_r$ refined from the Brillouin fit. (c) $t_{2g}$ and $e_g$ crystal field levels of octahedral $Co^{3+}$ and $Co^{2+}$ in low, intermediate and high spin states (LS, IS, HS), with expected ranges of the effective moment taken from Refs. [51,64,65] and tetrahedral $Co^{2+}$ shown for comparison. 3*d* orbital energy splitting caused by the Jahn-Teller effect is not shown.

Given that $Co^{2+}$ is evidently present, the question then becomes whether a tetrahedral-to-octahedral coordination transition is induced by the 1D steric confinement. Since this is not clear from the magnetization, the structural models must be reconsidered: if one replaces the $[Co^{II}Cl_4]^{2-}$ tetrahedra with $[Co^{II}Cl_6]^{4-}$ octahedra, the same general ADF motifs and intensities are reproduced in simulations (see Fig. S7) since only the chlorine positions change which are at the limit of detection. However, the stoichiometry of the unit cell changes to $[Cs_2Co^{II}Cl_5]^-$, violating the charge neutrality condition and implying that the encapsulated structure would need to donate a very significant amount of charge (1 $e^-$ per unit cell) to the SWCNT in order to be stable, additionally being contrary to the general behaviour of metal halides inside SWCNTs *i.e.* charge-neutral or acting as mild electron acceptors [28,58,59,67] akin to chromium chloride or the tetragonal polymorph described in Fig. 1. Therefore, it is concluded that a tetrahedral crystal field of $Co^{2+}$ is the most likely scenario, and even in the hypothetical case of $[Cs_2Co^{II}Cl_5]^-$ a slightly less massive orthorhombic strain $\varepsilon = 0.169$ is still necessarily generated in order to maintain congruence with the imaged motifs. From the similarly small Weiss constants of the bulk and confined samples ($-7.9(3)$ vs $-4.09(7)$ K respectively), the abundant $\not{p}mcm$ 1D phase is expected to order antiferromagnetically at similarly low temperatures to the strongly frustrated bulk precursor $Cs_2CoCl_4$ (~200 mK), with no significant renormalization of the net interactions.

In conclusion, transmission electron microscopy analysis and multislice simulations reveal that $Cs_2CoCl_4$ crystallizes in at least two one-dimensional polymorphs within single-walled carbon nanotubes (SWCNTs), specifically in the tetragonal rod group $\not{p}4/mcc$ and orthorhombic rod group $\not{p}mcm$. The formation of both structures is dependent on the SWCNT diameter (~2.6 and ~1.3 nm respectively), with the latter hosting a massive re-entrant orthorhombic strain of 25.8%. Extensive knowledge of the sample constraints permitted quantitation of the global SWCNT filling ratio and effective moment from Brillouin and Curie-Weiss fits to the measured d.c. magnetization (52(10)% and 4.788(3) $\mu_B$/f.u. respectively), confirming the persistence of $Co^{2+}$ in contrast with the well-known electron-doping of similar halides within SWCNTs. Comparison of simulations suggest that a tetrahedral $Co^{2+}$-centred crystal field is most plausible, with octahedral crystal field symmetry necessitating one-electron transfer per unit cell from the crystal to the SWCNT in order to meet the charge neutrality condition. Our findings show that the charge-lattice coupling in $Cs_2CoCl_4$ under 1D structural confinement topologically preserves the core magnetic properties of the bulk in contrast to the prevailing paradigm for metal halides, despite the drastic changes in the global structural symmetry. The advent of recent groundbreaking techniques in the electron microscope such as ptychography, 4D-STEM and exit wave reconstruction offers promising avenues in future analysis of precise atomic displacements and strain in these 1D heterostructures, which may be decomposed in terms of the irreducible representations of the subperiodic groups to construct a symmetry-based formalism of the charge-spin-lattice coupling, greatly accelerating exploration of the unknown phase space.

## ACKNOWLEDGEMENTS

We acknowledge use of the University of Warwick's Electron Microscopy and X-Ray Diffraction Research Technology Platforms (RTPs) in this work, and use of the magnetometers in the Department of Physics. Part of the HRTEM investigations performed at IP-PAS by K.B. and J.S. was supported by the National Science Centre (Poland) under Project No. UMO-2021/41/B/ST5/02588. We also acknowledge use of the instruments and scientific and technical assistance at the Monash Centre for Electron Microscopy (Monash University), a Microscopy Australia (ROR: 042mm0k03) facility supported by NCRIS. This research used equipment funded by Australian Research Council grant LE170100118. Data for Fig. S1(a) were collected using block allocation group time (reference CY39378-2) at Diamond Light Source I11. Patrick Ruddy is thanked for assistance in ampoule sealing.

## DATA AVAILABILITY

Microscope images, simulations, structural models of the bulk phase and modelled nanostructures (as .cif and CrystalMaker® .cmdx files respectively), and magnetization data are all available on the Warwick Research Archive Portal (WRAP) under reference number 202002.

---

# Supplementary information for 'Structural properties of one-dimensional $Cs_2CoCl_4$ confined within single-walled carbon nanotubes'

Jaskaran S. Mangat,[1] Yu Lei,[2] Matthew Weyland,[3,4] Yisong Han,[2] Kiran Bal,[2] Martin R. Lees,[2] Craig I. Hiley,[1] Piotr Dłużewski,[5] Sławomir Kret,[5] Peng Wang,[2] Richard I. Walton [1] and Jeremy Sloan [2]

*[1]Department of Chemistry, University of Warwick, Gibbet Hill Road, Coventry, CV4 7AL, United Kingdom.*
*[2]Department of Physics, University of Warwick, Gibbet Hill Road, Coventry, CV4 7AL, United Kingdom.*
*[3]Monash Centre for Electron Microscopy, Monash University, Clayton, Victoria, 3800, Australia.*
*[4]Deparment of Materials Science and Engineering, Monash University, Clayton, Victoria, 3800, Australia.*
*[5]Instytut Fizyki PAN, Al. Lotników 32/46, 02-668 Warszawa, Poland.*

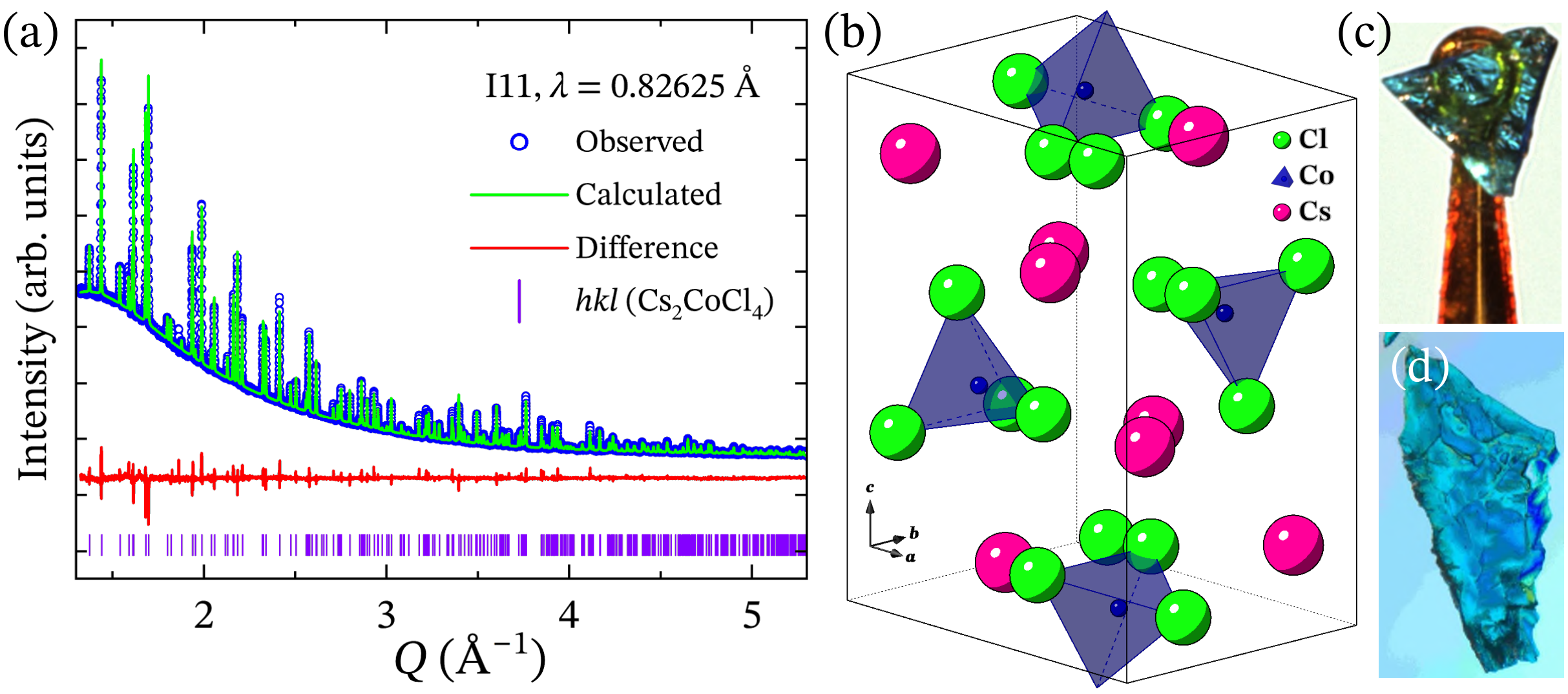

FIG. S1. (a) Rietveld refinement against powder X-ray diffraction data of $Cs_2CoCl_4$ within a 0.2 mm diameter borosilicate capillary. (b) Crystal structure of $Cs_2CoCl_4$ (two $Cl^-$ ions which are part of the tetrahedra are excluded since they are outside of the unit cell). (c) Twinned crystal of dimensions $0.304 \times 0.254 \times 0.041$ mm$^3$ used for the single crystal diffraction experiment (95.01 : 4.99 twin fractions). (d) A larger crystal of $Cs_2CoCl_4$.

TABLE S1. Single-crystal refinement of the twinned crystal shown in Fig. S1(c). The highest $U_{iso}$ values for each element were used for clTEM simulations of the nanostructures. Refinement was performed up to a fourth-order anharmonic potential to account for residual electron density observed in the difference Fourier map (similarly observed previously by Bagautdinov *et al.* for the chemical relative $Cs_2HgCl_4$ [1]), but for the simulations only the diagonal $\mathbf{U}_{ij}$ matrix elements were necessary to estimate $U_{iso}$ for the nanostructures. Structural information can be found in the attached .cif file (merged with .hkl and .res files).

| $Cs_2CoCl_4$ | #62 *Pnma* | $a$ / Å = 9.7776(2) | $b$ / Å = 7.4005(1) | $c$ / Å = 12.9850(2) | $V$ / Å$^3$ = 939.58(3) |
|---|---|---|---|---|---|
| Statistics | 85240 *hkl* collected | GoF on $F^2$ = 1.084 | $R_1$ ($I \geq 2\sigma(I)$) = 0.0189 | $wR_2$ = 0.0517 | Largest diff. peak/hole $e$·Å$^{-3}$ = 0.62/−0.66 |
| Site | $x$ | $y$ | $z$ | $U_{iso}$ / Å$^2$ | Occupancy |
| Cs01 | 0.52064(2) | 0.25 | 0.676097(17) | 0.03218(10) | 1 |
| Cs02 | 0.14074(3) | 0.75 | 0.60091(3) | 0.04806(12) | 1 |
| Co03 | 0.73435(4) | 0.75 | 0.57799(3) | 0.02455(11) | 1 |
| Cl04 | 0.81282(12) | 0.75 | 0.41402(8) | 0.0447(3) | 1 |
| Cl05 | 0.50421(10) | 0.75 | 0.59642(10) | 0.0468(3) | 1 |
| Cl06 | 0.82254(9) | 0.50260(11) | 0.65410(8) | 0.0516(2) | 1 |

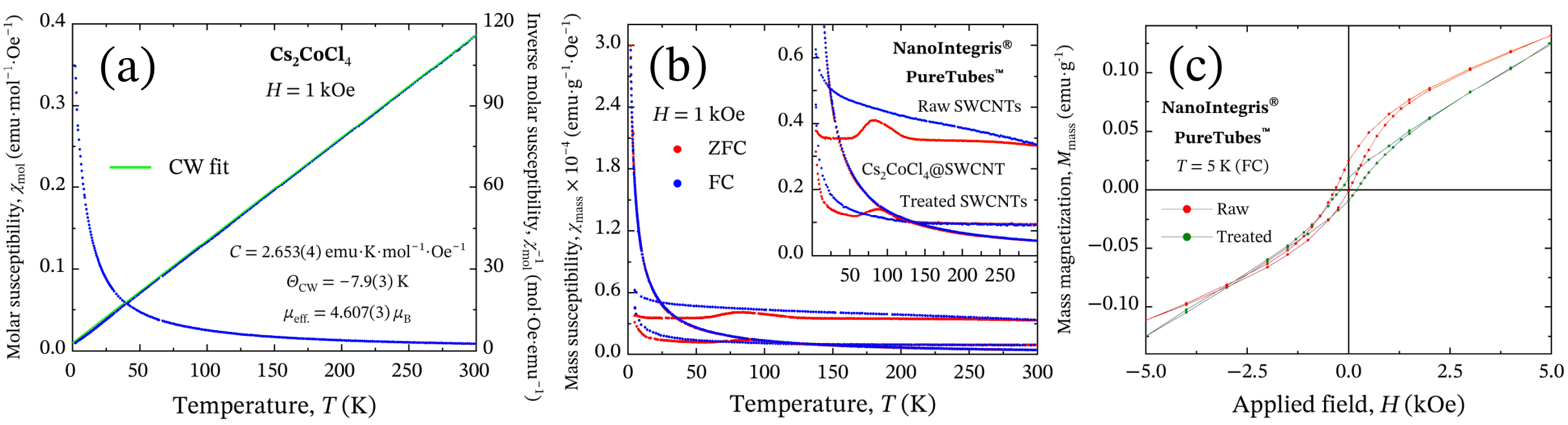

FIG. S2. (a) Field-cooled (FC) Curie-Weiss fit for the powdered precursor $Cs_2CoCl_4$. (b) Comparison of the temperature-dependent susceptibilities of raw NanoIntegris® PureTubes™ SWCNTs (single-walled carbon nanotubes), the $Cs_2CoCl_4$@SWCNT sample after washing, and treated NanoIntegris® PureTubes™ SWCNTs (same washing protocol as $Cs_2CoCl_4$@SWCNT). The susceptibility of the SWCNTs after washing is lowered significantly, indicating that most magnetic catalyst (generally Ni and Fe, both ferromagnetic at room temperature) was successfully removed. The $Cs_2CoCl_4$@SWCNT sample has an even lower high-temperature susceptibility with no stark transitions, suggesting the presence of even less catalyst and indicating that the magnetization of the SWCNTs is somewhat batch-dependent. The hump in the zero-field cooled (ZFC) magnetization is most likely caused by $O_2$. (c) Field-dependent magnetization for both raw and treated SWCNTs. The saturation in the high-field regime is strongly neutered after washing.

| Element | At% |
|---|---|
| Cs | 33.5 |
| Co | 10.5 |
| Cl | 55.9 |

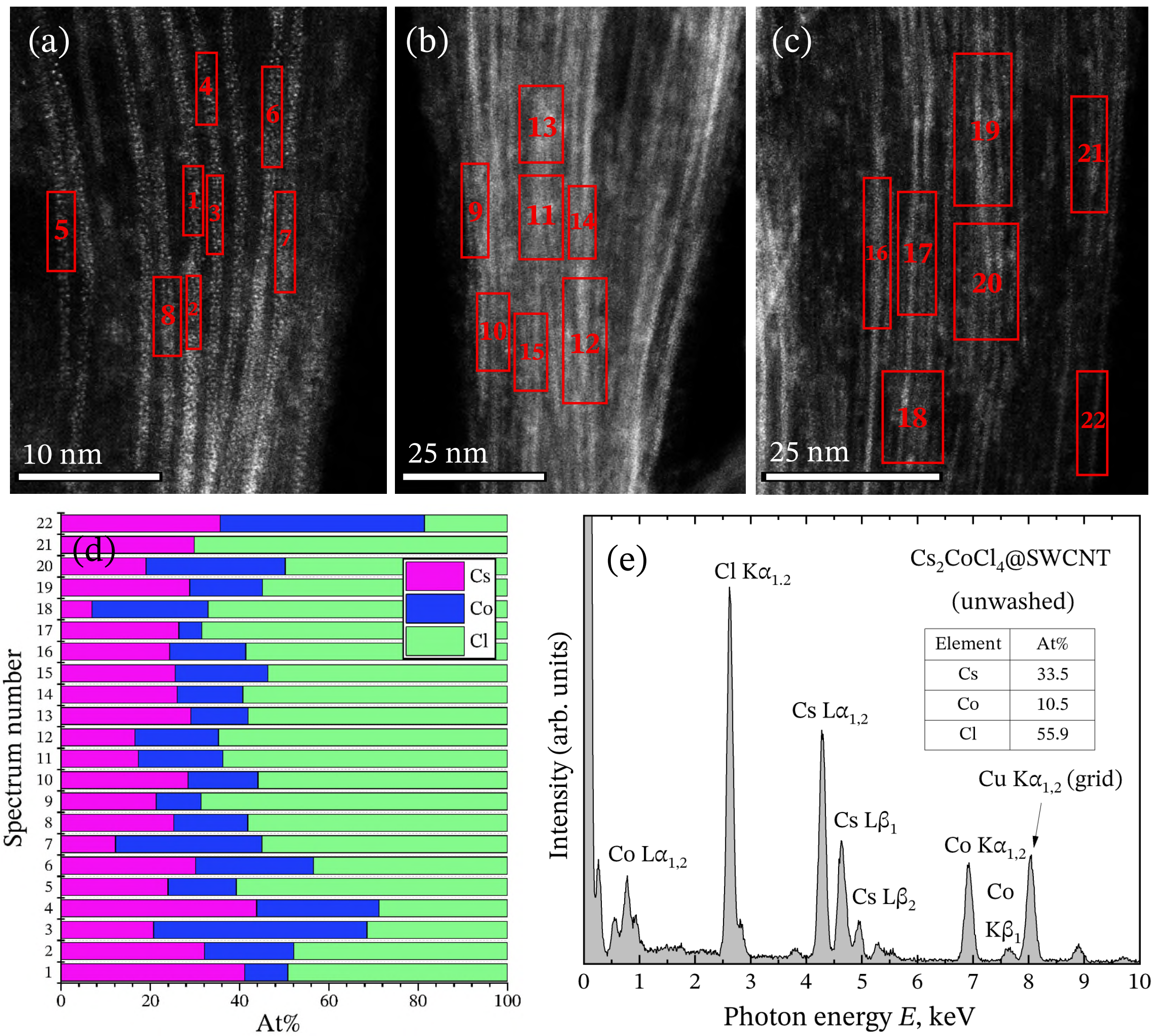

FIG. S3. (a, b, c) Areal regions used for energy-dispersive X-ray spectroscopy (EDX) scans of nanotube bundles in the final washed sample. (d) Atomic percentage of Cs:Co:Cl across the areal regions. For comparison, the nominal percentage ratio of $Cs_2CoCl_4$@SWCNT is 28.6 : 14.3 : 57.1 (disregarding carbon). (e) Prototypical EDX spectrum before washing.

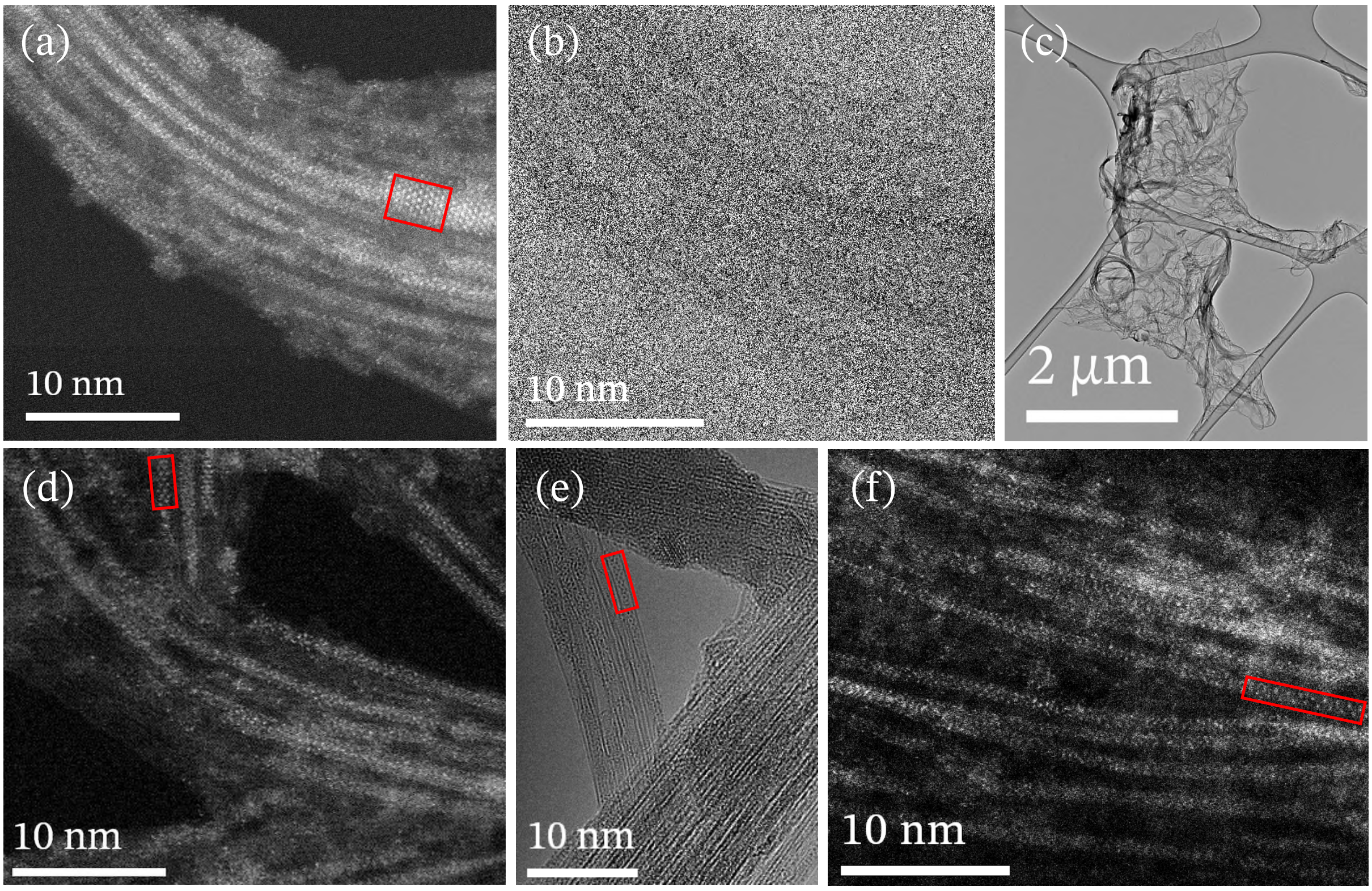

FIG. S4. (a, b) Annular dark-field (ADF) and bright-field (BF) images of the region around the structure analysed in Fig. 1, showing the heavy presence of carbon near the right side. (c) Finely dispersed SWCNT bundles in the final sample (*i.e.* post-washing). (d, e) ADF and high-resolution transmission electron microscopy (HRTEM) images of the regions around the hexagonal motifs in Fig. 2. (f) ADF region around the zigzag motif of Fig. 3.

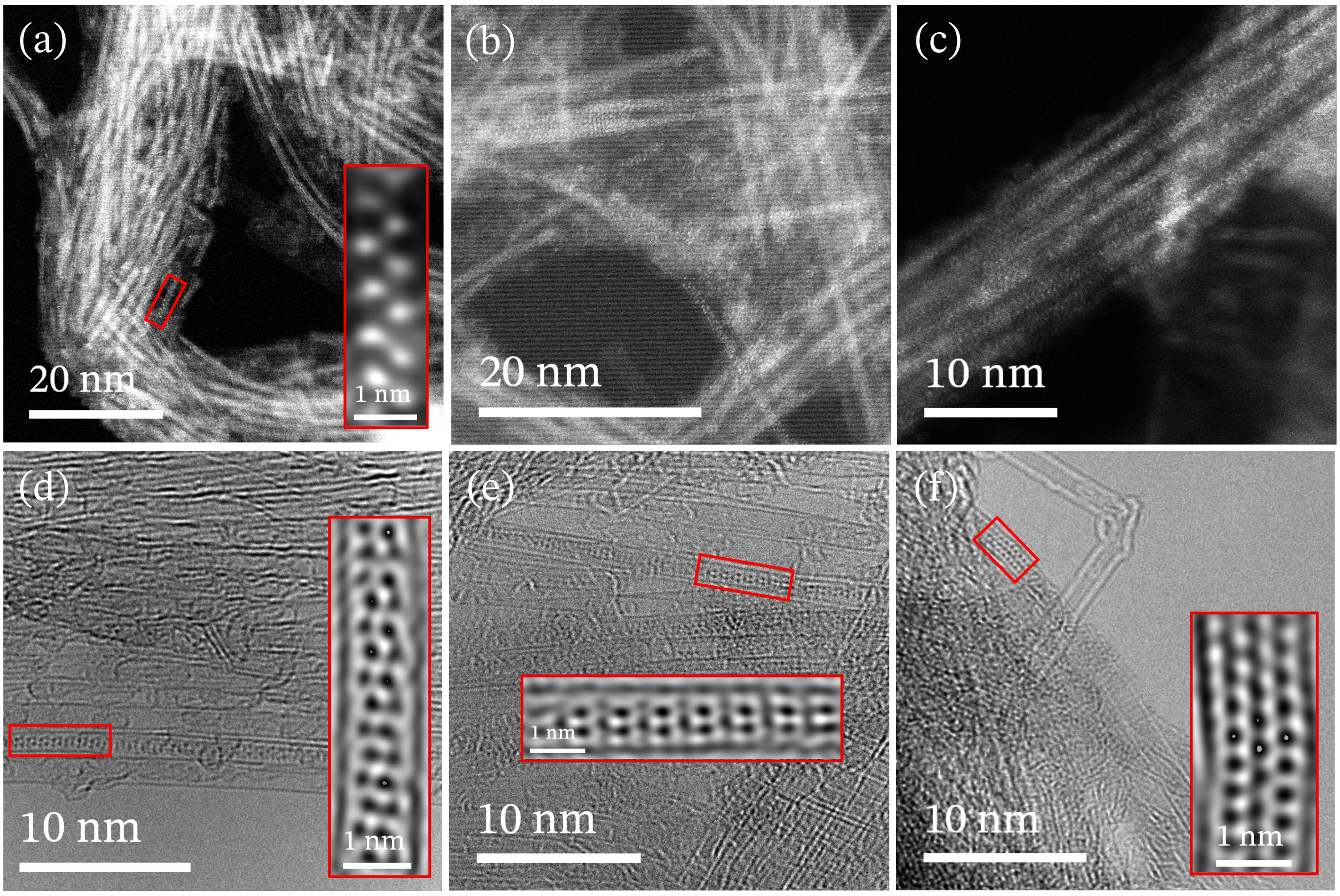

FIG. S5. (a) ADF image of a high filling ratio region in the SWCNT bundles, containing another 'zigzag' example of $Cs_2CoCl_4$@SWCNT, similar to that depicted in Fig. 3. (b, c) Examples of orientationally-randomized and orientationally-preferred filled SWCNT bundles. (d, e) Additional HRTEM images of $Cs_2CoCl_4$@SWCNT exhibiting the hexagonal atomic motif initially depicted in Fig. 2. (f) Example of a Cs-rich confined crystal as a sample inhomogeneity.

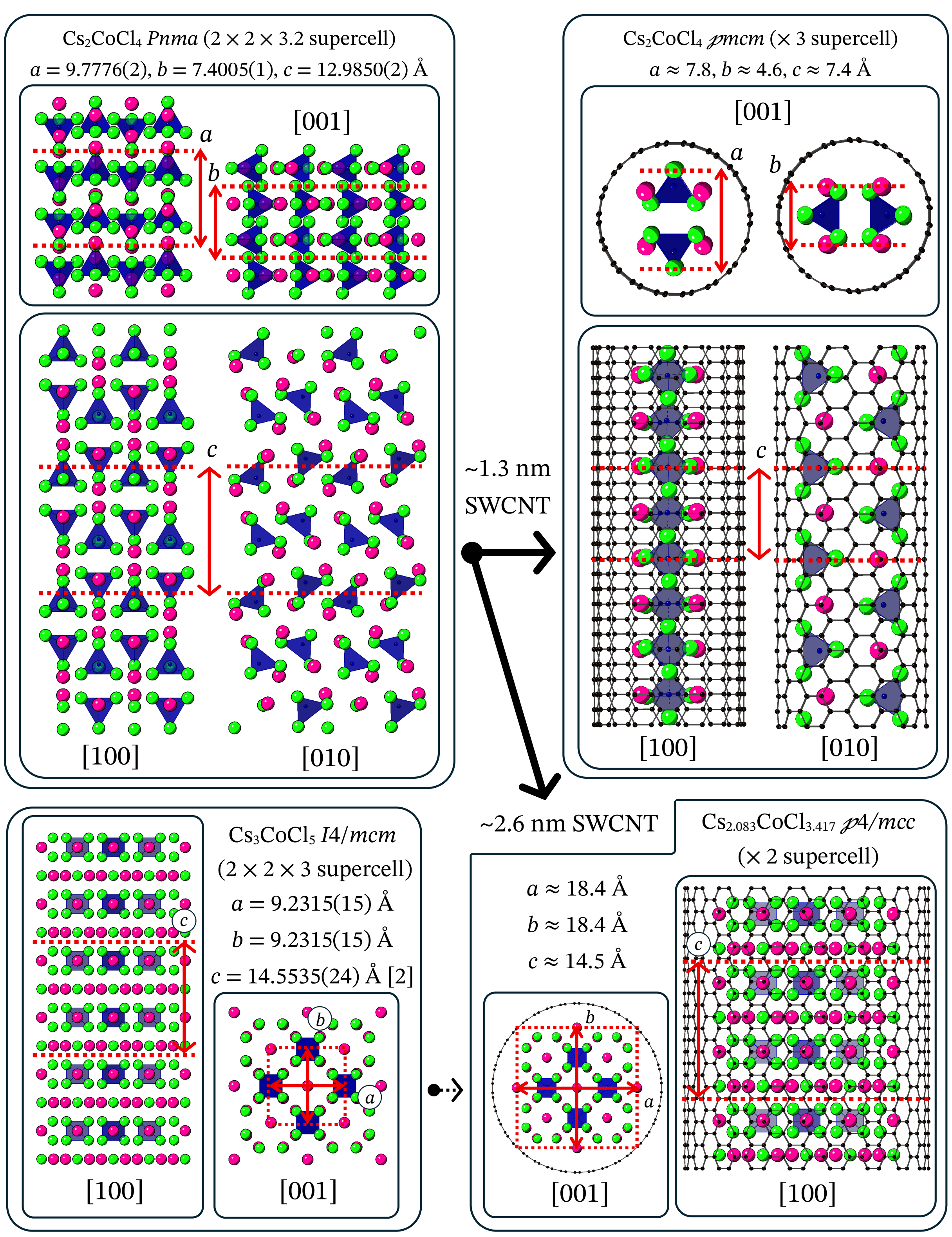

FIG. S6. Structural map showing the relationships between the bulk and confined structures of $Cs_2CoCl_4$. The [001] perspective clearly shows inequal dimensions (orthorhombicity) in the basal plane after confinement within ~1.3 nm wide SWCNTs whilst maintaining the ternary chemical composition of the precursor; whereas in ~2.6 nm wide SWCNTs the structure is essentially the same as $Cs_3CoCl_5$, but with four columns of CsCl at the cross-sectional diagonal extrema removed and thereby necessitating a change in the stoichiometry (approximating $Cs_2CoCl_{3.4}$).

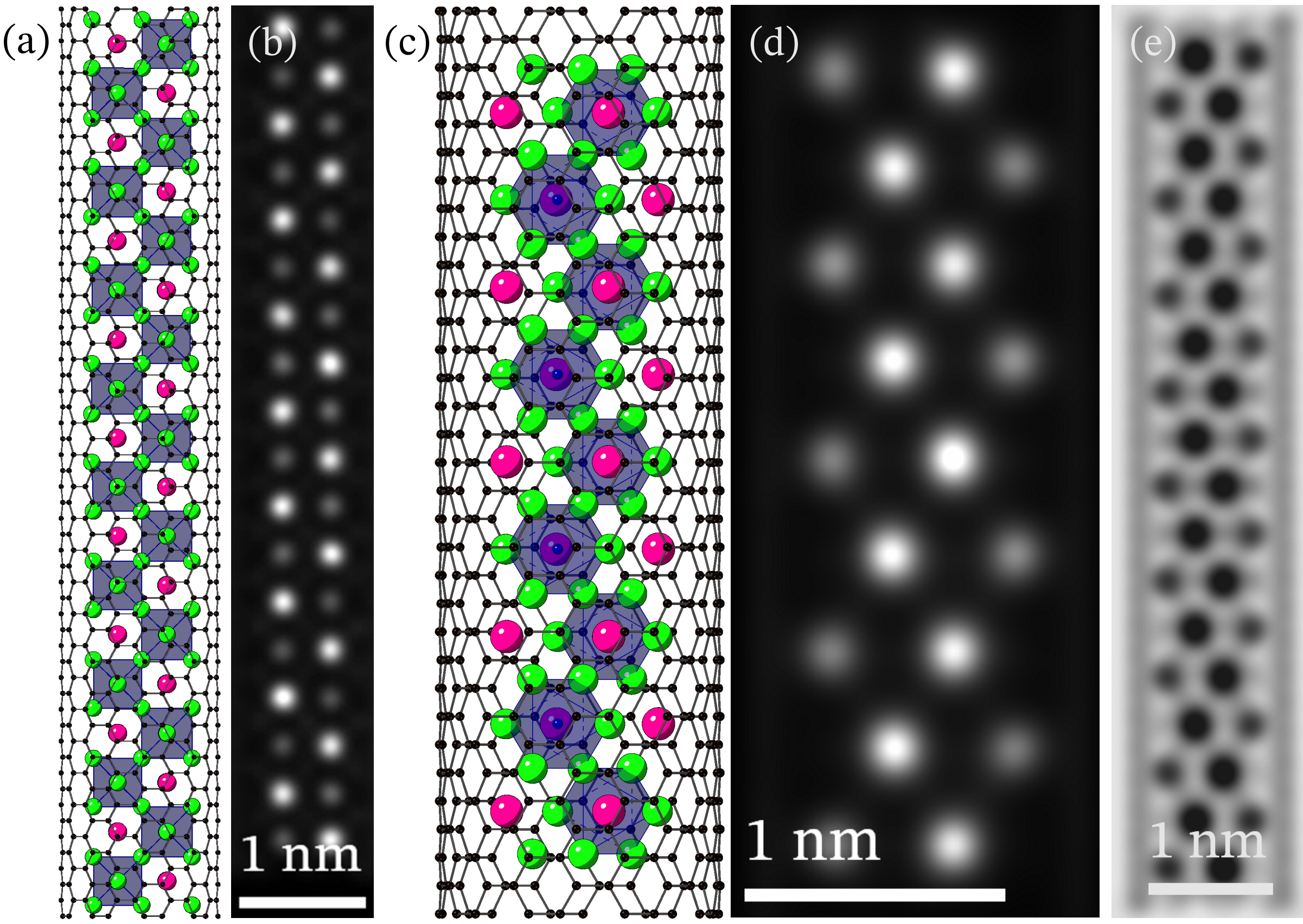

FIG. S7. (a, b) Model and ADF simulation of an alternative *pmcm* orthorhombic structure (lattice parameters $a \approx 7.6$ Å, $b \approx 5.4$ Å and $c \approx 7.6$ Å) in the same zigzag orientation as Fig. 3, but hosting corner-shared $Co^{2+}$-centred octahedra instead of isolated tetrahedra. The atomic periodicities and $Cs_2$-dominated incoherent scattering are very similar to the tetrahedral model, however, the $CoCl_2$ scattering in the voids between the $Cs_2$ is significantly more point-like (less blurred) than the tetrahedral and experimental cases in Fig. 3. By inspection the unit cell necessitates an empirical formula of $Cs_2CoCl_5$, implying 1 $e^-$ charge transfer from the crystal to the SWCNT due to our d.c. magnetization analysis constraining the oxidation state of cobalt to be $Co^{2+}$. (c, d, e) 55° rotation of the model in a near-identical orientation to Fig. 2, with accompanying ADF and HRTEM simulations.

---